# Inertia and feedback parameters adaptive control of virtual synchronous generator

Hai-Peng Ren, Qi Chen, Liang-Liang Zhang, Jie Li
Xi'an University of Technology, Xi'an, China, 710048

**Abstract**: The virtual synchronous generator technology analogs the characteristics of the synchronous generator via the controller design. It improved the stability of the grid systems which include the new energy. At the same time, according to the adjustable characteristics of the virtual synchronous generator parameters, the parameter adaptive adjustment is used to improve the dynamic performance of the system. However, the traditional virtual synchronous generator adaptive control technology still has two drawbacks: on the one hand, the large-scale adjustment of the damping droop coefficient and the virtual moment of inertia requires the system having a high energy storage margin; On the other hand, there is a power overshoot phenomenon in the transient regulation process, which is disadvantageous to the power equipment. First, this paper provides a convenient adjustment method for improving the transient stability of the system, the system damping is adjusted by introducing the output speed feedback. Second, according to the transient power-angle characteristics of the system, a parameter adaptive control strategy is proposed, which shortens the transient adjustment time and ensures that the deviation of the system frequency in the transient adjustment process is within the allowable range, and improves the transient performance of the grid frequency adjustment, at the same time, the power overshoot is suppressed. Finally, the experimental results show that the proposed control strategy is superior to the existing adaptive control strategy.

## 1 Introduction

With the development of renewable energy generation, more and more renewable energy, such as solar and wind power, is connected to the power grid through the power electronic converters. Most renewable energy has intermediate generation volume, which makes the grid be subject to the instability [1]. To deal with this situation, virtual synchronous generator (VSG) was proposed to provide the converter based renewable energy with the synchronous generator external property, which provides an effective tool for stability regulation of the power grid with high renewable energy penetration rate [1-4].

In the traditional VSG technique, there are four parameters needed to be tuned, including the virtual inertia $J$, the frequency drooping coefficient $D_p$, the integral gain $K$ and the voltage droop coefficient $D_p$, among them, $D_p$, and $D_p$ are determined by the national regulation [5-6], which mean the active power change caused by the frequency vibration and the reactive power change caused by the voltage amplitude variation [3], respectively. The VSG transient stability referred to as the ability to maintain frequency stable when transient perturbation occurs. The traditional methods maintain and improve the transient stability by regulating $D_p$ and $J$. Reference [8,9] proposed $J$ adaptive regulation method based on the analysis of the transient state of the power versus angle characteristics, which was proved to be stable using Lyapunov theory. However, the adaptive expression $J$ was not given in these references. Reference [10] gave the J selection principle by using the small signal model analysis of the distributed power system. Reference [11] determined the J adaptive regulation method using fuzzy logic reasoning. Above mentioned methods considered the function of virtual inertia parameter $J$ in the stability improvement, without considering the function of the damping parameter $D_p$ on the transient stability of the system.

References [13,14] analyzed the contribution of the damping parameter Dp to the transient stability of the VSG system, and proposed a parameter adaptive method using the damping parameter, Dp, according to the frequency fluctuation. This method was proved to effectively reduce the overshoot of the active power with the reduced settle down time. Reference [15] adaptively regulated the damping parameter, Dp, according to the active power deviation feedback. References [16-19] analyzed the contribution of J and Dp to the transient stability and obtained the selection principle of J and Dp to enhance the transient stability of the VSG system. On the base of the work in [16-19], reference [20] proposed a particle swarm optimization method using the frequency deviation and voltage deviation as index function to achieve the optimal J and Dp, where the stability of the system was proved by using Lyapunov direct method. Reference [21] established the transient method of VSG, and proposed a $J$ and $Dp$ joint adaptive regulation method considering comprehensively the transient indices including overshoot and settle down time, etc.. Different from the above mentioned methods, Reference [22] proposed the optimal angle acceleration criterion on the basis of the analysis the oscillation of the active power and frequency in the transient state.

To conclude the above references, there exist three parameter adaptive strategies to enhance the transient performance of the VSG, including J adaptive regulation, Dp adaptive regulation and J/Dp joint adaptive regulation. The idea of these methods is adaptively regulating the J and/or Dp to improve the transient performance of the VSG system according the analysis of the power versus angle characteristics of the VSG system. These method show two main disadvantages: first, there exists large transient overshoot of the active power, which leads to the large transient voltage and current pressure to the devices in the converter, thus brings threaten to the reliability operation of the converters; second, to achieve the satisfactory transient performance, J and Dp are adjusted in a large range, which requires the large storage redundancy of the system, at the same time, large J means the large lag in the power frequency regulation loop, and weak stability the system. Meanwhile, large Dp adjustment violates the regulation of the power grid running rules although it is done only in the transient state.

In this paper, to solve the aforementioned problems, a output speed feedback is proposed to regulate the system damping, and thus provide a tool to regulate the system transient state in order to avoid the Dp regulation in the transient state. It is more important that the method avoids large J regulation and alleviates the large storage redundancy in the VSG. The experimental results show the effectiveness and superiority of the proposed method.

The rest of the paper is arranged as follows: The basic principle of VSG is given, and the damping regulation mechanism of the proposed output speed feedback is demonstrated in Section 2. The parameters adaptation method is proposed in the Section 3 to enhance the transient performance of the VSG system. The experimental results and comparisons are given in Section 4 to show the effectiveness and superiority of the proposed method. The conclusions are given in Section 5.

## 2 Basic principle of VSG

VSG working in a grid connection mode can be treated as an ideal voltage source serial with an output impedance connected to the grid as shown in Fig. 1.

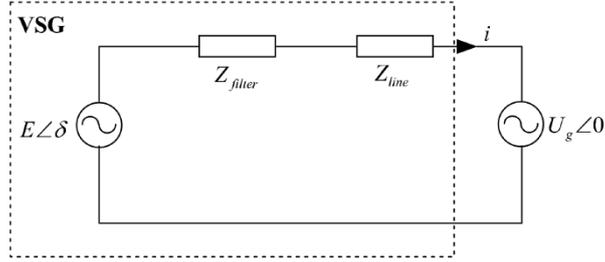

Fig.1 VSG equivalent model

In Fig.1, $E$ is the three-phase inverter generated voltage; $U_g$ is the grid phase voltage, $Z_{filter}$ is the filter impedance; $Z_{line}$ is the line impedance; $i$ is the current fed into the grid; $\delta$ is the power angle, which represents the phase difference between the three phase converter voltage and the grid voltage, as given by

$$\delta = \int(\omega - \omega_0)dt ,  \tag{1}$$

where $\omega_0$ is the fundamental angular frequency of the grid; $\omega$ is the virtual synchronous angular frequency of the converter, $\delta$ is the power angle.

The output current of the VSG can be given as

$$\dot{I} = \frac{E\angle\delta - U_g\angle 0}{Z} = \frac{E\angle\delta - U_g\angle 0}{r + jX} \tag{2}$$

where $Z$ include two parts, i.e., $Z_{line}$ and $Z_{filter}$, $r$ represents resistance, $X$ represents inductance, thus, $\alpha = \tan^{-1}(X/r)$ is defined as Impedance angle. Here, the apparent power of the VSG is

$$\begin{aligned} S &= 3U_g \dot{I}^* = 3U_g \frac{E\angle(-\delta) - U_g\angle 0}{r - jX} \\ &= 3\frac{U_g}{Z}(E\cos(\alpha - \delta) - U_g\cos\alpha + jE\sin(\alpha - \delta) - jU_g\sin\alpha) \\ &= P_e + jQ \end{aligned} \tag{3}$$

where * represents conjugate.

From (3), we have the active power out $P_e$ and reactive power output $Q_e$ given by

$$\begin{cases} P_e = 3\dfrac{EU_g}{Z}\cos(\alpha - \delta) - 3\dfrac{U_g^2}{Z}\cos\alpha \\ Q_e = 3\dfrac{EU_g}{Z}\sin(\alpha - \delta) - 3\dfrac{U_g^2}{Z}\sin\alpha \end{cases} \tag{4}$$

For an inductance line and small power angle, we have

$$\begin{cases} P_e = 3\dfrac{EU_g}{Z}\delta \\ Q_e = 3\dfrac{EU_g - U_g^2}{Z} \end{cases} \quad (5)$$

VSG animates the synchronous generator by using swing equation given by

$$J\omega_0 \frac{d\omega}{dt} = P_m - P_e - D(\omega - \omega_0) \quad (6)$$

where $\dfrac{d\omega}{dt}$ is virtual angle acceleration, $P_m$ is the VSG input power, $P_e$ is the VSG output power, $D = \omega_0 D_p$ is damping. The small signal model can be used to obtain the transform function from power angle to the output power given by

$$H_{p\delta}(s) = 3\frac{EU_g}{Z} \quad (7)$$

To this end, we have the block diagram of the active power control loop of VSG system as shown in Fig.2.

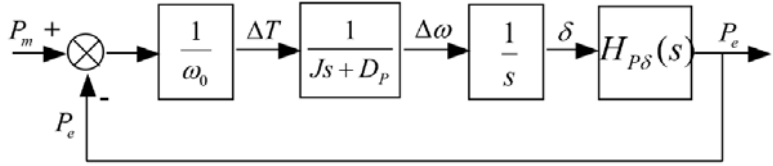

Fig.2 VSG active power control loop diagram

From Fig.2, we have the open loop transform function of the active power given by

$$G(s) = \frac{P_e}{P_m} = \frac{H_{p\delta}(s)}{J\omega_0 s^2 + D_p \omega_0 s} \quad (8)$$

From Eq. (8), we have the damping ratio of the system as

$$\zeta = \frac{D_p}{2}\sqrt{\frac{\omega_0}{J}}\sqrt{\frac{Z}{3EU_g}} \quad (9)$$

From Eq. (9), The system damping ration can be adjusted by $D_p$ and $J$, however, $D_p$ and $J$ adjustment is restricted by storage volume. Meanwhile, if the J is very large, then it is subjected to the power angle oscillation [24]. To solve this problem, in this paper, we introduce the output speed feedback to provide a tool for damping ratio adjustment in order to improve the VSG system transient stability. The system with output speed feedback is given in Fig.3.

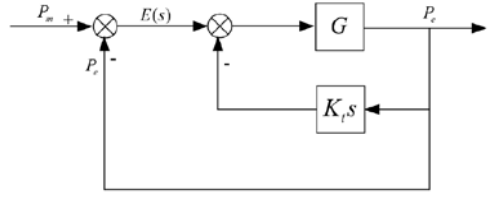

Fig. 3 Block diagram of output speed feedback control

From Fig.3, we have the close loop transfer function of the system given by

$$\phi(s) = \frac{G(s)}{1+G(s)(1+K_t s)}$$
$$= \frac{H_{P\delta}(s)}{J\omega_0 s^2 + (D_p\omega_0 + H_{P\delta}(s)K_t)s + H_{P\delta}(s)} \quad (10)$$
$$= \frac{H_{P\delta}(s)/J\omega_0}{s^2 + \dfrac{D_p\omega_0 + H_{P\delta}(s)K_t}{J\omega_0} s + H_{P\delta}(s)/J\omega_0}$$

where is the active power open loop transfer function given by Eq. (8). From Eq. (10), we know that the natural frequency $\omega_n$ and damping ratio $\zeta$ of system (10) are

$$\begin{cases} \omega_n = \sqrt{H_{P\delta}(s)/(J\omega_0)} \\ \zeta = \dfrac{D_p\omega_0 + H_{P\delta}(s)K_t}{2\sqrt{H_{P\delta}(s)\times J\omega_0}} \end{cases} \quad (11)$$

It can be seen from (11), the system damping ratio can be adjust by extra parameter $K_t$. it can be given by

$$K_t = \frac{2\zeta\sqrt{H_{P\delta}(s)\times J\omega_0} - D_p\omega_0}{H_{P\delta}(s)} \quad (12)$$

## 3 parameters adaptive control of VSG

### 3.1 The relationship between *J* and the transient stability of the system

In this paper, the method of simultaneous adjustment of $J$ and $K_t$ is used to avoid the need for large energy storage reserves and the change of $D_p$ in the dynamic process of adjustment. The swing equation of VSG after introducing speed feedback can be expressed as:

$$J\frac{d\omega}{dt} = \frac{P_m - P_e - K_t \dfrac{dP_e}{dt}}{\omega_0} - D_p(\omega - \omega_0) \quad (13)$$

It can be transformed into:

$$\frac{d\omega}{dt} = \frac{P_m - P_e - K_t \dfrac{dP_e}{dt} - D_p\omega_0(\omega - \omega_0)}{J\omega_0} \quad (14)$$

The virtual angular velocity change rate dω/dt is inversely proportional to the virtual moment of inertia $J$. That is, the larger $J$ is, the slower the virtual angular velocity ω changes. Conversely, the smaller $J$ is, the faster ω changes. The next section will design the parameter adaptation law according to the relationship between the rate of change of angular velocity and the virtual moment of inertia described in equation (14).

**3.2 Analysis of transient response process of VSG disturbance**

When the power disturbance breaks the power balance, according to the synchronization mechanism of "active power adjusting frequency", the VSG autonomously adjusts the input power $P_m$ to restore the system balance. The transient adjustment process diagram is shown in Figure 5. Figure 5(a) shows the relationship between VSG power angle and input power, Figure 5(b) is a schematic diagram of frequency change curve during adjustment.

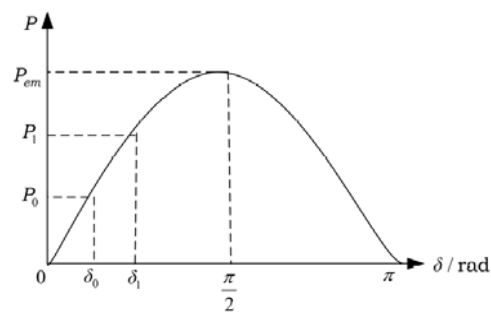

(a) VSG power versus angle curve

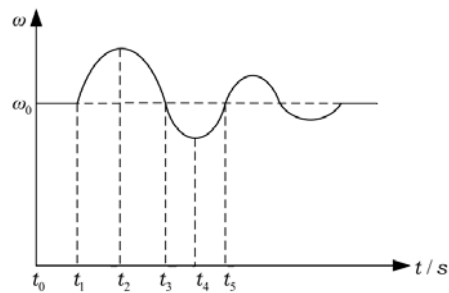

(b) Frequency variation during the transient stage

Fig.5 VSG power angle curve and frequency variation

The detail analysis of the VSG transient adjustment process and its power angle characteristics is as follows:

$t_0 - t_1$ stage: VSG input active power $P_m = P_0$, which is basically equal to the output electromagnetic power $P_e$, that is, $P_0 \approx P_e$, ω ≈ ω0, dω/dt ≈ 0, and the system is in a balanced state.

$t_1 - t_2$ stage: The power disturbance causes the input power $P_m$ to jump to $P_1$, while the electromagnetic power $P_e$ remains instantaneously unchanged, resulting in power imbalance, that is, $P_m = P_1 > P_e = P_0$. At this time, Δω = (ω − ω $_0$)> 0, dω/dt> 0, it can be seen that, at this stage, the frequency deviation is positive, and at the same time, the frequency change rate is also positive, which means that the frequency deviation is changing in the increasing direction.

If the absolute value of the frequency change rate is also large at this time, reducing the frequency change rate by increasing $J$ to suppress the increase in frequency (and frequency deviation).

Therefore, the regulation law of $J$ is:

$$J = \begin{cases} J_0 + \dfrac{k_1}{e^{\left|\frac{\Delta\omega}{2\pi}\right|}}, & \left|\dfrac{d\omega}{dt}\right| \geq T \\ J_0, & \left|\dfrac{d\omega}{dt}\right| < T \end{cases} \quad (15)$$

Where: $J_0$ is the initial value of the virtual moment of inertia: $\Delta\omega/2\pi = \Delta f$ is the degree of deviation of the system frequency from the grid frequency; $T$ is the virtual angular velocity change rate threshold set to prevent frequent adjustment of $J$; $k_1$ is the adjustment coefficient, It can be calculated according to

$$k_1 = (J_{max} - J_0)e^{|\Delta f_{max}|} \quad (16)$$

where: $\Delta f_{max}$ is the allowable threshold for system frequency changes, according to the national standard GB/T15945-2008 ``Power Quality Power System Frequency Allowable Deviation'': under the normal operation of the power system, the frequency deviation range is ±0.2Hz. If the installed capacity of the microgrid system is small, the margin can be enlarged, and the allowable frequency deviation value is 0.5 Hz. This article sets $\Delta|f_{max}|$ = 0.5 Hz, and the setting of $k_1$ in equation (16) can ensure that the maximum of $J$ is $J_{max}$.

The greater the value of the virtual moment of inertia $J$, the worse the dynamic performance of the system, and the greater the phase angle lag caused at the cut-off frequency, which leads to the reduction of the phase angle margin of the system, and may even cause the continuous oscillation of the VSG power angle. This article adds output speed feedback control, which can achieve strong frequency adjustment ability without excessive $J$ adjustment. Therefore, the setting of $J_{max}$ deviating from $J_0$ can be much smaller than simply using virtual moment of inertia adjustment. Therefore, the amount of change in energy storage capacity is reduced.

$t_2 - t_3$ stage: $\Delta\omega > 0$, $d\omega/dt < 0$, the virtual angular velocity $\omega$ is in a decelerating state. At this time, it is not necessary to make adjustments and keep the frequency naturally close to the desired value. If the absolute value of the frequency change rate during the approach is too large, then the number of oscillations will increase. Therefore, when the absolute value of the frequency change rate is too large, the virtual moment of inertia $J$ is reduced, so that the negative (less than 0) angular velocity change rate increases in the increasing direction (also the direction in which the absolute value decreases) approaching 0.

Therefore, the regulation law of $J$ is:

$$J = \begin{cases} J_0 - \dfrac{k_2}{e^{\left|\frac{\Delta\omega}{2\pi}\right|}}, & \left|\dfrac{d\omega}{dt}\right| \geq T \\ J_0, & \left|\dfrac{d\omega}{dt}\right| < T \end{cases} \quad (17)$$

$k_2$ is the adjustment coefficient, and its value is determined by equation (18).

$$k_2 = (J_0 - J_{min})e^{|\Delta f_{max}|} \quad (18)$$

$J_{min}$ should not deviate too far from $J_0$, because too much deviation from $J_0$ will result in a larger angular velocity change rate hindering frequency stability. Due to the output speed feedback effect, the adjustment of *J* can be small, and the amount of change corresponding to the energy storage capacity is correspondingly small. The amount of energy storage change with respect to the nominal virtual moment of inertia is reduced.

$t_3 - t_4$ stage: Δω <0, dω/dt< 0, it is similar to the $t_1 - t_2$ stage, the virtual angular velocity ω is in the acceleration stage. The adjustment method of *J* is the same as that in the $t_1 - t_2$ stage.

$t_4 - t_5$ stage: Δω <0, dω/dt> 0, it is similar to the case of $t_2 - t_3$ stage, the adjustment method of *J* is the same as that of the $t_2 - t_3$ stage.

The following part analyzes the damping parameter adaptation law. In the transient adjustment process, in order to ensure that the system power does not appear obvious overshoot, at the same time, to shorten the transient adjustment time, we set the system damping ζ = 1.1. The corresponding speed feedback coefficient $K_t$ can be calculated by (7), (12). But the following two situations should be noted:

The first case: When the system frequency change is greater than the defined frequency threshold, that is |Δf|> Δfmax = 0.5Hz, in order to suppress the frequency increase, set the frequency change rate dω/dt = 0, according to formula (14), it can be calculated:

$$K_t = \frac{P_m - P_e - \omega_0 D_p(\omega - \omega_0)}{\dfrac{dP_e}{dt}} \tag{19}$$

The second case: When the system frequency change is within the safety threshold, that is |Δf| <= Δfmax = 0.5Hz, | dω/dt |> *T* indicates that the frequency fluctuates seriously, the speed feedback coefficient is used to increase the system damping, we choose ζ = 1.3 to stabilize the system frequency as soon as possible.

In summary, the VSG parameter adaptive control law proposed in this paper is as follows:

If Δω <2πΔf max:

$$J = \begin{cases} J_0 + \dfrac{k_1}{e^{\left|\frac{\Delta\omega}{2\pi}\right|}}, & \Delta\omega\dfrac{d\omega}{dt} > 0 \text{ and } \left|\dfrac{d\omega}{dt}\right| > T \\ J_0 - \dfrac{k_2}{e^{\left|\frac{\Delta\omega}{2\pi}\right|}}, & \Delta\omega\dfrac{d\omega}{dt} < 0 \text{ and } \left|\dfrac{d\omega}{dt}\right| > T \\ J_0, & others \end{cases} \tag{20}$$

$K_t$ can be obtained by equations (7), (12), where the system damping is selected as equation (21).

$$\zeta = \begin{cases} 1.3 & \left|\dfrac{d\omega}{dt}\right| > T \\ 1.1 & \left|\dfrac{d\omega}{dt}\right| \leq T \end{cases} \tag{21}$$

Otherwise, if Δω> 2πΔf, then the law is

$$\begin{cases} J = J_0 \\ K_t = \dfrac{P_m - P_e - \omega_0 D_p(\omega - \omega_0)}{\dfrac{dP_e}{dt}} \end{cases} \tag{22}$$

After the above-mentioned stages of regulation, the system power reaches a new equilibrium state.

### 3.3 VSG parameter adaptive stability analysis

After adding the output speed feedback control, the closed-loop transfer function of the system is shown in equation (10), and the characteristic root of the system can be obtained as:

$$s_{1,2} = \frac{-A \pm \sqrt{B}}{2J\omega_0} \quad (23)$$

Where, $A = D_p\omega_0 + \frac{3EU_g}{Z}K_t$; $B = (D_p\omega_0 + \frac{3EU_g}{Z}K_t)^2 - 4J\omega_0\frac{3EU_g}{Z}$.

Because $J\omega_0 > 0$, the distribution of $s_{1,2}$ in the s-plane is mainly determined by the numerator. If the system is stable, the parameter $A > 0$, that is, $D_p\omega_0 + \frac{3EU_g}{Z}K_t > 0$, we get:

$$K_t > -\frac{D_p\omega_0}{\frac{3EU_g}{Z}} \quad (24)$$

On this basis:

1). When $B < 0$, $s_{1,2} = \frac{-A \pm \sqrt{B}}{2J\omega_0}$ is a pair of conjugate complex roots in the left half plane, and the system is stable;

2). When $0 \leq B < A^2$, $s_{1,2} = \frac{-A \pm \sqrt{B}}{2J\omega_0}$ is the negative real root on the left half plane, and the system is stable;

3). When $B \geq A^2$, $s_1 = \frac{-A - \sqrt{B}}{2J\omega_0} < 0$ is on the left half plane, $s_2 = \frac{-A + \sqrt{B}}{2J\omega_0} > 0$ is on the right half plane, then the system is unstable.

In summary, the conditions that need to be met for ensuring the system stability are:

$$\begin{cases} A > 0 \\ B < A^2 \end{cases} \quad (25)$$

That is,

$$\begin{cases} K_t > -\dfrac{D_p\omega_0}{\dfrac{3EU_g}{Z}} \\ J > 0 \end{cases} \quad (26)$$

Obviously, $J>0$ can always be guaranteed in the adaptive control process, and the value of $K_t$ can

be analyzed as follows:

a) When $\Delta\omega \leq 2\pi\Delta f_{max}$, $K_t = \dfrac{2\zeta\sqrt{J\omega_0 \dfrac{3EU_g}{Z}} - D_p\omega_0}{\dfrac{3EU_g}{Z}} = \dfrac{2\zeta\sqrt{J\omega_0 \dfrac{3EU_g}{Z}}}{\dfrac{3EU_g}{Z}} + \dfrac{-D_p\omega_0}{\dfrac{3EU_g}{Z}} > \dfrac{-D_p\omega_0}{\dfrac{3EU_g}{Z}}$, meet the stability condition;

b) When $\Delta\omega > 2\pi\Delta f_{max}$, $K_t = \dfrac{P_m - P_e - D_p\omega_0(\omega - \omega_0)}{\dfrac{dP_e}{dt}}$.

According to the definition of the damping droop coefficient $D_p$ [2,3]:

$$D_p = \frac{\Delta P}{\omega_0(\omega - \omega_0)} = \frac{P_e - P_{e0}}{\omega_0(\omega - \omega_0)} \tag{27}$$

Where: $P_{e0}$ represents the VSG power before disturbance.

According to Eq (1) and Eq (5):

$$\frac{dP_e}{dt} = 3\frac{EU_g}{Z}(\omega - \omega_0) + 3\frac{U_g}{Z}\delta\frac{dE}{dt} \tag{28}$$

If the system stability conditions are met, the following inequality should be satisfied

$$K_t = \frac{P_m - P_e - \omega_0 D_p(\omega - \omega_0)}{\dfrac{dP_e}{dt}} > -\frac{D_p\omega_0}{\dfrac{3EU_g}{Z}} \tag{29}$$

Move the right side of the above inequality to the left side and substitute Eq (27) into it, we get:

$$\frac{\dfrac{3EU_g}{Z}(\omega - \omega_0)(P_m - 2P_e + P_{e0}) + \dfrac{dP_e}{dt}(P_e + P_{e0})}{\dfrac{3EU_g}{Z}\dfrac{dP_{e0}}{dt}(\omega - \omega_0)} > 0 \tag{30}$$

This situation is the control strategy when the system frequency fluctuates beyond the limit. Other than this case, there are two situations. The first situation is that the system frequency exceeds the rated threshold in the positive direction, in this situation the following inequalities are satisfied:

$$\begin{cases} \omega - \omega_0 > 0 \\ \dfrac{dP_e}{dt} > 0 \\ P_m - P_e > 0 \end{cases} \tag{31}$$

From Eq (31), we can see that the denominator of equation (30) is greater than 0, so inequality (30) only needs to have its numerator polynomial greater than 0. Substituting Eq (28) into Eq(30) numerator polynomial, the condition of obtaining numerator polynomial greater than zero is:

$$\frac{3EU_g}{Z}(\omega - \omega_0)(P_m - P_e) + \frac{3EU_g}{Z}(\omega - \omega_0)(P_{e0} - P_e) + \frac{3EU_g}{Z}(\omega - \omega_0)(P_e - P_{e0}) + \frac{3U_g}{Z}\delta\frac{dE}{dt}(P_e - P_{e0}) > 0 \tag{32}$$

The second and third terms of the above equation can be cancelled, so the stable conditions is:

$$\frac{3EU_g}{Z}(\omega-\omega_0)(P_m-P_e)+\frac{3U_g}{Z}\delta\frac{dE}{dt}(P_e-P_{e0})>0 \qquad (33)$$

because $E$ is a slow physical quantity, d$E$/d$t$ is very small and approximates to 0, so the above formula becomes:

$$\frac{3EU_g}{Z}(\omega-\omega_0)(P_m-P_e)>0 \qquad (34)$$

It can be seen from Eq (31) that the above inequality is always true, that is, the system is stable.

In another case, when the frequency is less than the rated value and the limit is exceeded, all "greater" signs in equation (31) become "less" signs. Similar deductions can lead to the same conclusion. So far, the stability of the parameter adaptive adjustment process has been proved.

Compared with the traditional constant damping and constant virtual moment of inertia control, the system cannot instantaneously change from the initial point to the new equilibrium point. When the acceleration area exceeds the deceleration area, transient instability occur [7]. According to the response characteristics of different phases in the system transient state, an adaptive control is a real-time control strategy, which not only improves the transient stability of the system, enhances the robustness of the system, at the same time, improves the dynamic performance of the system. The characteristics of the adaptive control strategy proposed in this paper are: On the one hand, the output speed feedback control is proposed so that the large-range adjustment of the damping droop coefficient $D_p$ and the virtual moment of inertia $J$ is no longer needed, which suppresses the system power overshoot. At the same time, the transient adjustment time is shortened. On the other hand, the adaptive adjustment of the speed feedback coefficient limits the system frequency variation within the threshold during the dynamic process, which effectively avoids the VSG from being disconnected from the grid due to the large frequency change during the transient stage.

## 4 Experimental validation

In order to verify the correctness of the above theoretical analysis, a three-phase inverter prototype is built based on the TMS320F28335 digital controller. The inverter output is connected to the grid through a transformer. The parameters of the prototype are shown in Table 1. The switch tube uses the IGBT module FF150R12RT4, the switch tube drive chip is based on 1EDI60I12AF, the inductor current signal on the inverter side is sampled by the current hall CSNE151-100, and the capacitor voltage and grid voltage signals are sampled by the voltage hall HNV025A. The experimental setup is shown in Figure 6. The experiment control sampling period is 200μs, and the inverter switching frequency is 10kHz.

The initial active power and reactive power of the VSG are 157W and 0Var, respectively. Active power steps from 157W to 600W at 6s, while reactive power maintains 0Var. The experimental results are shown in Figures 7-9. Among them, $J$ adaptive control is the method of literature [8-9]; $D$p adaptive control is the method of literature [13]; $J$ and $D$p parameter adaptive control is the method of literature [16 ].

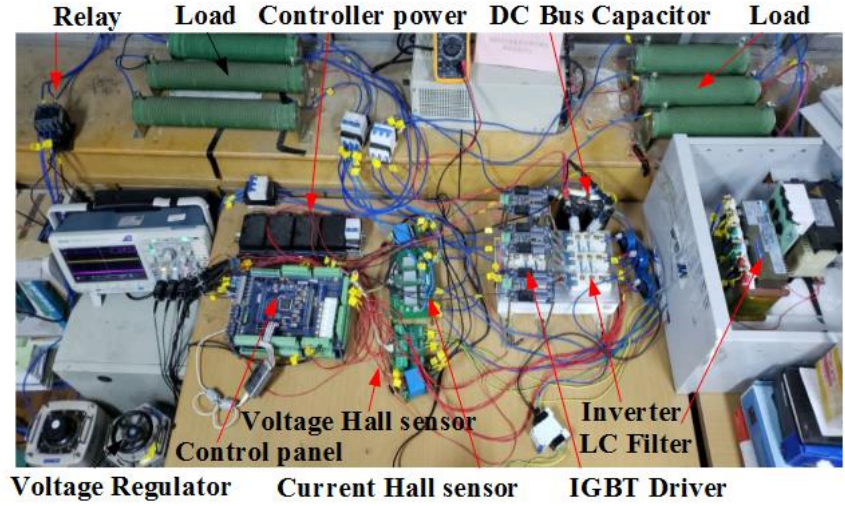

Fig.6 Experiment platform

Table 1 The main parameters

| Parameter | Value |
|---|---|
| Input voltage Vin/V | 250 |
| Grid voltage(rms) Vg/V | 70.7107 |
| Inverter-side inductor L/H | 7*10-3 |
| Filter capacitor C/F | 4*10-6 |
| Grid-side inductor $L_{line}$/H | 2*10-3 |
| Grid-side parasitic resistance $R_{line}/\Omega$ | 0.6 |
| The initial value of the virtual moment of inertia J | 0.0025 |
| Integral gain K | 2000 |
| Mechanical friction coefficient $D_p$ | 0.3 |
| Voltage droop coefficient Dq | 42.4264 |

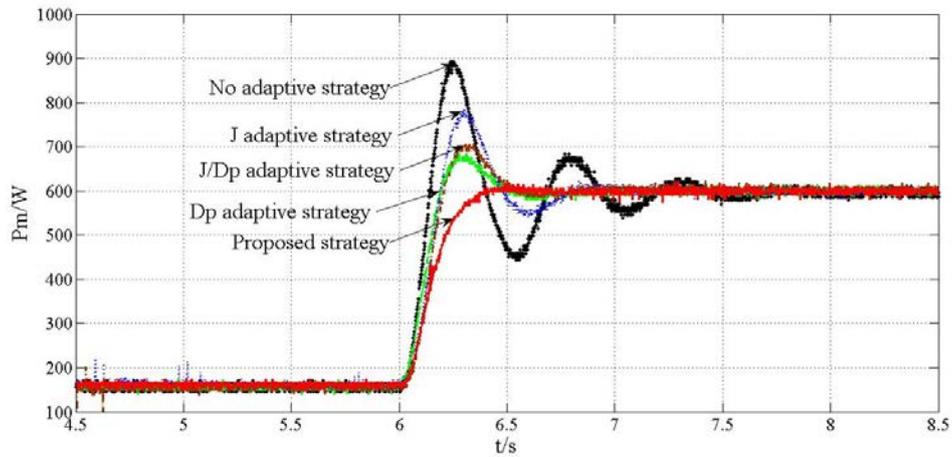

Fig.7 VSG power responses of different control methods

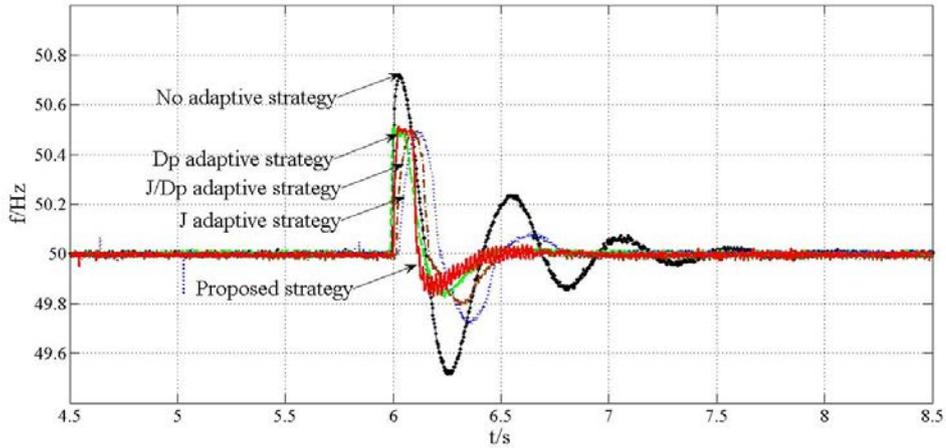

Fig.8 VSG frequency responses of different control methods

Figures 7 and 8 reflect the transient power response and frequency response curves of the method in this article and other existing methods during the adjustment process. From the experiment results, it can be seen that, using the method of constant $J/D_p$, the power overshoot in the transient process is as high as 50%, the adjustment time is 1.5s, and the system frequency instantaneously exceeds the safe working limit, i.e., 0.5 Hz, causing the VSG to go offline. With $J$ adaptive control, system frequency fluctuation and power overshoot are reduced to a certain extent, but the power overshoot is still obvious. At the same time, it can be seen from Figure 9(a) that if $J$ is used to adjust separately, the adjustment range of $J$ is larger; Using $D_p$ adaptive control, when and only when the system frequency exceeds the threshold, increase the damping droop coefficient $D_p$ to limit frequency fluctuations. Figure 9 (b) shows that the rapid $D_p$ vibration corresponding; With simultaneous adaptive adjustment of $J$ and $D_p$, the performance has been improved, but there are still the above problems. Compared with the method in this paper, its dynamic response time is long, and the instantaneous adjustment of $J/D_p$ is also larger. Using the control method proposed in this paper, the system works in an over-damped state, and the power overshoot is suppressed. At the same time, the system frequency fluctuation range is limited within the range of 0.5 Hz, and the system can be quickly restored to a stable state, which improves the system transient process.

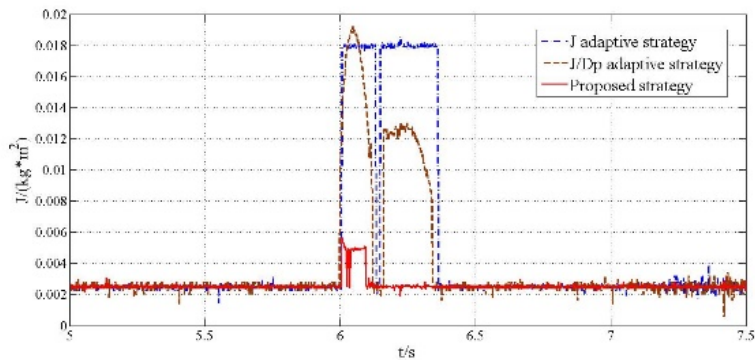

(a) $J$ adjustment of different methods

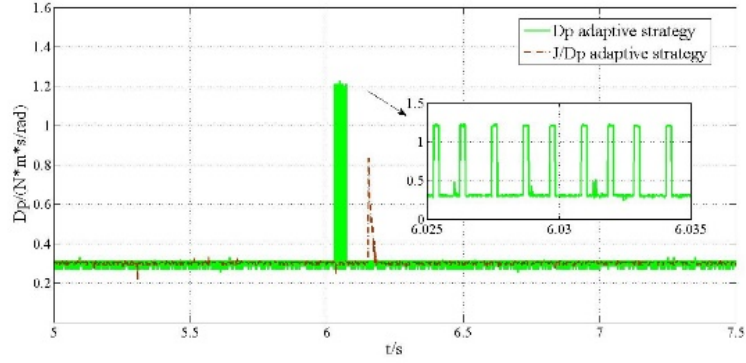

(b) $D_p$ adjustment of different methods

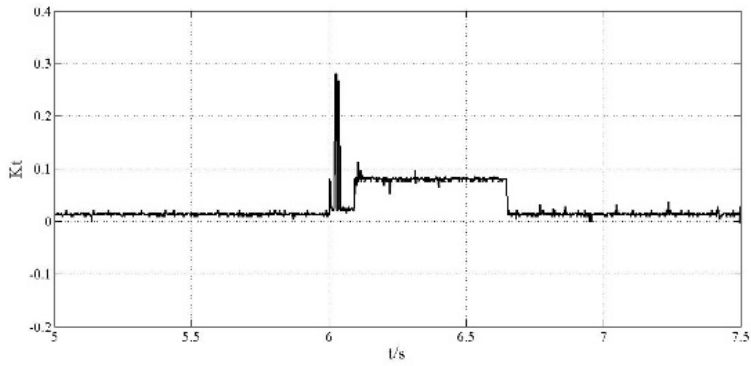

(c) $K_t$ adjustment of different methods

Figure 9 shows the parameter adjustment curves of several methods compared through experiments, in which subgraph (a) is the change curve of $J$ corresponding to the three methods that need to adjust $J$, subgraph (b) is $D_p$ adjustment, and $D_p$ change curve in $J/D_p$ adjustment, sub-figure (c) is the curve of the output speed feedback coefficient $K_t$ using the method in this paper. From the subgraph (a), in order to limit the frequency threshold, the maximum virtual moment of inertia that the J adaptive control strategy needs to provide is 0.019, which is 7.6 times the initial value; The maximum virtual moment of inertia required by the $J/D_p$ adaptive control strategy is also as high as 0.018; while the instantaneous virtual moment of inertia required by the method in this article is only 0.0056, it can be seen that the adaptive control method in this article requires the smallest virtual moment of inertia adjustment. The adjustment time of the proposed method is significantly shortened. It can be seen from the subgraph (b) that when $D_p$ adaptive control or $J/D_p$ adaptive control strategy are adopted, the required damping droop coefficient varies widely, and the maximum instantaneous value reaches 4 and 2.8 times, respectively, with respect to the steady state value ($D_{p0}$ = 0.3), which requires higher energy storage capacity redundancy ; It can be seen from the subgraph (c) that the system damping is maintained at ζ = 1.1 at the beginning, and the change trend of the speed feedback coefficient $K_t$ is consistent with the change trend of $J$ in the method in the subplot (a). When the system frequency exceeds 0.5 Hz, $J$ decreases, $K_t$ increases, and the system damping is increased according to Eq (12), which limits the frequency within the threshold. Subsequently, the higher $K_t$ value effectively reduces the frequency deviation, and finally, when the system enters the steady state, $K_t$ returns to the initial value.

Based on the above experimental results, comparing the existing three different parameter adaptive adjustment methods, the following conclusions are obtained:

1) Using $J$ adaptive adjustment, in order to stabilize the dynamic process, the instantaneous value of $J$ required is large, which leads to a decrease in system damping, thus the system works in an under-damped state, and power overshoot occurs, as seen experimental results are shown in Figures 7 and 9 (a) .

2) Using $D_p$ adaptive adjustment, increasing $D_p$ can increase system damping, but the increase of $D_p$ only occurs when the frequency exceeds the set value. Once the frequency drops within the set value, the initial value is used. The adjustment process is as follows as shown in Figure 9(b), the power response also has overshoot, as shown in Figure 7.

3) Using $J/D_p$ adaptive adjustment, the experimental results are shown in Figure 9 (a) and (b). During the transient adjustment process, $J$ and $D_p$ are adjusted at the same time, which cannot effectively ensure that the system is in an over-damped state, which leads to the system overshoot, as shown by the power response results in Figure 7.

The advantage of the method in this paper is that it suppresses power overshoot, limits the transient frequency within the threshold, and has excellent response speed.

## 5 Conclusions

VSG technology enables renewable energy power generation systems that use power electronic converters to connect to the grid to independently participate in the primary frequency regulation of the grid, which improves the frequency stability of the system. The existing VSG adaptive control strategy use the virtual moment of inertia or the damping droop coefficient for adjustment thus requiring large energy storage redundancy. However, the existing methods still have the problems of large parameter adjustment range, long adjustment time, large transient overshoot, and easily cause the VSG to run off the grid. To solve these problems, this paper proposes a VSG adaptive control strategy based on the output speed feedback. The main contributions are as follows:

1) Using output speed feedback control to adjust the damping of the system makes the system work under the over-damping characteristic, avoiding frequent repeated charging and discharging of energy storage equipment, and at the same time, avoiding the power overshoot. During the adjustment process, the adjustment range of the virtual moment of inertia is limited, so that the dynamic adjustment performance of the VSG can be improved without excessive energy storage capacity redundancy.

2) Based on the analysis of the VSG power angle characteristics curve and the transient adjustment process, a new VSG adaptive control principle is obtained using the feedback gain and virtual inertia, which can suppress the power overshoot in the dynamic process, speed up the adjustment process, and limit the frequency fluctuation range. It is ensured that the VSG will not be disconnected due to frequency overrun during the dynamic process. Meanwhile, it does not need the large virtual inertia change.